\documentclass{sig-alternate}
\usepackage{amsfonts}
\usepackage{listings}
\usepackage{amsmath}
\usepackage{amssymb}
\usepackage{graphicx}
\newcommand{\ignore}[1]{}

\begin{document}

\title{Fast Data in the Era of Big Data: Twitter's\\ Real-Time Related Query Suggestion Architecture}

\author{
{Gilad Mishne, Jeff Dalton, Zhenghua Li, Aneesh Sharma, Jimmy Lin}
\vspace{1.6mm}\\
Twitter, Inc.\\
@gilad @jeffd @zhenghuali @aneeshs @lintool
}

\date{}
\maketitle

\begin{abstract}
We present the architecture behind Twitter's real-time related query
suggestion and spelling correction service. Although these tasks have
received much attention in the web search literature, the Twitter
context introduces a real-time ``twist'':\ after significant breaking
news events, we aim to provide relevant results within minutes.  This
paper provides a case study illustrating the challenges of real-time
data processing in the era of ``big data''. We tell the story of how
our system was built twice:\ our first implementation was built on a
typical Hadoop-based analytics stack, but was later replaced because
it did not meet the latency requirements necessary to generate
meaningful real-time results. The second implementation, which is the
system deployed in production, is a custom in-memory processing engine
specifically designed for the task. This experience taught us that the
current typical usage of Hadoop as a ``big data'' platform, while
great for experimentation, is not well suited to low-latency
processing, and points the way to future work on data analytics
platforms that can handle ``big'' as well as ``fast'' data.
\end{abstract}

\hyphenation{Map-Reduce}

\section{Introduction}

According to a well-known clich\'{e}, there are three aspects of
data:\ {\it volume}, {\it velocity}, and {\it variety}. To date,
most of the focus, both in academia and industry, has been on volume,
although many realize the importance of velocity
(exemplified by work on streaming and online
algorithms~\cite{Aggarwal_2007} and recent open-source projects such
as Storm\footnote{\small
  http://storm-project.net/}) as well as variety (exemplified by
attempts at integrating structured, semi-structured, unstructured, and
even multimedia data). This paper tells the story of how we tried to
deal with velocity with an architecture designed for volume, learned
about the limitations of the approach, and completely rewrote the
system to handle ``fast data''. The process has been instructive, and
we wish to share our designs and the lessons learned with the
community.

The context of this work is related query suggestion and spelling
correction in search, which we collectively called ``search
assistance'' at Twitter.\footnote{\small
  http://engineering.twitter.com/2012/05/related-queries-and-spelling.html}
Although both problems have been studied in detail in the web
context~\cite{Cui_etal_2003,Cucerzan_Brill_EMNLP2004,JonesRosie_etal_WWW2006,Mei_etal_CIKM2008},
Twitter brings a new ``twist'' to the problem:\ search assistance needs
to be provided in real time and must dynamically adapt to the rapidly
evolving ``global conversation''. What exactly do we mean by ``real
time''?  In providing search assistance, we need to balance
accumulating sufficient evidence to return relevant results with
responsiveness to newly-formed associations between queries that often
derive from breaking news stories.  For example, prior to Marissa
Mayer's appointment as Yahoo's CEO, the query ``marissa mayer'' had
little semantic connection to the query ``yahoo''; but following news
of that appointment, the connection is immediate and obvious.  Based
on our study of how rapidly Twitter queries evolve to reflect
users' interests, we aim for a target latency of ten minutes. That is,
ten minutes after a major news event breaks, the service should be
displaying relevant related query suggestions. Delivering this level
of service requires us to tackle the ``velocity'' challenge mentioned
above.

Our first solution took advantage of Twitter's Hadoop-based analytics
stack built primarily around Pig and did not require building any
additional infrastructure. Since the analytics platform was
designed to handle petabyte-scale datasets through large batch jobs, it
proved to be ill-suited for the strict latency requirements of the
search assistance application. As a result, we abandoned our original
implementation and designed a completely different architecture
specifically for real-time processing. Throughout the process, we
gained a better understanding of how fundamental assumptions in
Hadoop's design make it a poor fit for real-time applications. We
detail the shortcomings of our initial Hadoop implementation and
describe how they are addressed in the actual deployed system. Note
that the focus of this work is on data processing architectures and
not the algorithms for computing related queries and spelling
corrections; the algorithms are discussed only to the extent necessary
to help the reader understand the architecture.

To be explicit, this paper is meant as a case study and not intended
to present novel research contributions. Nevertheless, we believe that
``war stories'' and practical experiences with building large-scale
data processing systems form a valuable part of the literature. We
view this paper as having three contributions:\

\begin{list}{\labelitemi}{\leftmargin=1em}
\setlength{\itemsep}{-2pt}

\item First, we introduce the {\it real-time} related query suggestion
  problem, attempt to define what ``real time'' actually means in this
  context, and articulate how it is different from similar problems in
  the web context.

\item Second, we describe two separate working systems that were built
  to solve the problem:\ the initial Hadoop-based implementation and
  the deployed in-memory processing engine. These experiences are
  valuable for understanding the limitations of Hadoop-based stacks.

\item Third, our experiences highlight a gap between architectures for
  processing ``big data'' and those for ``fast data''. We
  present thoughts on a future research direction for the field
  of data management in bridging these two worlds.

\end{list}

\noindent Organizationally, this paper makes the following
progression:\ after providing background, we present the first
iteration of the system (the Hadoop-based implementation). We then
describe the system written to replace it (the deployed system).
Finally, we discuss the need for a {\it general} and {\it unified}
data processing platform for ``big'' and ``fast'' data.

\section{Background}

We begin with a more detailed description of the problem and
challenges. Related query suggestion is a feature that most searchers
are likely familiar with:\ when the searcher types in a query (e.g.,
``obama''), in addition to showing results for the query, the system
suggests queries that might also be of interest (e.g., ``white
house''). Spelling correction can be viewed as a special case, where
the suggested query is a closely-related form of the
original:\ perhaps a missing or transposed character. For example,
``justin beiber'' is a common misspelling for ``justin bieber''.

\subsection{Related Work}

In information retrieval (IR), the general idea of augmenting a user's
query is closely related to relevance feedback, which dates back to
the 1960s~\cite{Rocchio_1965}. One specific form, {\it
  pseudo-relevance feedback}, automatically extracts expansion terms
from an initial query's top-ranked results
(see~\cite{Lavrenko_Croft_SIGIR2001} for a more modern
formulation). Whether the user controls the use of these additional
query terms is an interface design decision~\cite{Koenemann96}. We can
consider the case where expansion terms are explicitly controlled by
the user an early form of query suggestion---these and related
techniques have been widely known in the IR literature for decades and
predate the web.

Prior to the web, most query expansion work focused on capturing term
correlations across global and local contexts in the document
collection~\cite{Xu00}. The advent of web search engines, however,
provided a new and much richer resource to mine:\ query, clickthrough,
and other behavioral interaction logs. One of the earliest use of logs
for query expansion is the work of Cui et al.~\cite{Cui_etal_2003},
who used clickthrough data to establish correlations between query
terms and document terms, which were then extracted for query
expansion. Related, a family of query suggestion techniques involves
constructing a bipartite graph of query and clicked URLs, on which
random walks~\cite{Mei_etal_CIKM2008} or clustering can be
performed~\cite{CaoHuanhuan_etal_SIGKDD2008};
cf.~\cite{baraglia-time-flow-graph}. Another use of query logs is to
extract query substitutions from search sessions by mining statistical
associations from users' successive
queries~\cite{JonesRosie_etal_WWW2006}---this is the general approach
we adopt. Similar techniques are also effective for spelling
correction~\cite{Cucerzan_Brill_EMNLP2004}.

There has been much related work on analyzing temporal patterns of web
search queries. Vlachos et al.~\cite{vlachos-2004-bursts} were among
the first to model bursts in web queries to identify semantically
similar queries from the MSN query logs. The temporal profile of
queries has been analyzed~\cite{JonesRosie_Diaz_TOIS2007} and
exploited to capture lexical semantic
relationships~\cite{alfonseca-2009,shokouhi-season-queries}.
Forecasted query frequency has also been shown to be helpful in query
auto-completion~\cite{shokouhi-radinsky-2012}. Most recently, Radinsky
al.~\cite{radinsky-behavioral-dynamics-2012} proposed a general
temporal modeling framework for user behavior in terms of queries,
URLs, and clicks.

\subsection{Real-Time Related Query Suggestion}

We argue that the related query suggestion problem takes on
additional, richer dimensions in the Twitter context. A key
characteristic of Twitter is that it provides up-to-the-second updates
on major events around the world, ranging from Arab Spring protests to
the outcome of major sporting events to the sudden occurrences of
natural disasters. This means that related query suggestions must be
{\it real-time}:\ in particular, results need to be temporally
relevant and timely. We consider these two points in detail below.

In the information retrieval literature, relevance captures the notion
of the ``goodness'' of a result. It is a somewhat fuzzy notion, and IR
researchers have devoted countless pages over the past several decades
trying to more precisely define relevance~\cite{Mizzaro98}. Most
operational definitions of relevance focus on {\it topicality}, or the
``aboutness'' of a particular result. Despite a thread of work that
attempts to capture
temporality~\cite{LiXiaoyan_Croft_CIKM2003,Efron_Golovchinsky_etal_SIGIR2011,Dakka_etal_2012},
in the standard treatment, relevance is {\it atemporal}, i.e., merely
a function of the query and result, irrespective of {\it when} the
result was returned.  Applied to evaluate the relevance of related
query suggestions, we are not aware of any previous work that
explicitly attempts to factor in temporal issues.

It is clear that real-time related query suggestion has a strong temporal
component. Consider an example:\ on June 28, 2012, the hashtag
\mbox{\#SCOTUS}, short for Supreme Court of the United States, was
trending on Twitter, which indicates a large (and atypical) volume of
tweets on the topic. On that day, the Supreme Court delivered its
judgment on the constitutionality of President Obama's health care
reform. A click on the trend automatically triggers a query for the
hashtag:\ related query suggestions on that day included
``healthcare'' and ``\#aca'' (short for Affordable Care Act, the name
of the legislation). In this case, the service accurately captured the
connections between those keywords within the temporal context. The
same suggestions would have not been relevant a few days before, when
the Supreme Court was ruling on immigration legislation, and would not
be relevant some time later, when the court moves on to consider other
cases.

Another important difference between related query suggestion for web
search and the real-time variant of the problem is the narrow time
frame in which suggestions have maximal impact. Often, the temporal
progression of breaking news events on Twitter follows a ``hockey puck''
curve. When plotting, say, number of queries as a function of time, we
typically observe a region when the volume is increasing at a moderate
slope, followed by a transition to where the volume increases at an
accelerated rate (often exhibiting exponential growth). Ideally, we
would like to start making related query suggestions at the ``knee''
of that curve, which requires a delicate balancing act. If we make the
suggestion too early, the connections might be too tenuous due to
scant evidence. On the other hand, making suggestions too late would
lessen the impact, since users might have already found out about
related queries through other means. This issue of timeliness is less
important in the context of suggesting web queries (and as far as we
are aware, mostly ignored in the research literature).

To render the problem more challenging, rapid changes in the query
stream corresponding to breaking news events are intermixed with slower
moving signals that persist over longer periods of time:\ ``michelle
obama'' vs. ``flotus'' (First Lady of the United States) would be an
example. Furthermore, tail queries require accumulation of statistical
evidence across longer periods of time to make meaningful suggestions,
due to low query volume. An example might be plausibly making the
suggestion ``\#bigdata'' for ``hadoop''. Ideally, we desire a system
that deals with rapidly changing signals (high-volume, by definition)
as well as slowly changing signals, which may either be high or low
volume.

\subsection{Quantifying ``Churn''}
\label{section:bg:churn}

\begin{figure}[t]
\includegraphics[width=1.0\linewidth]{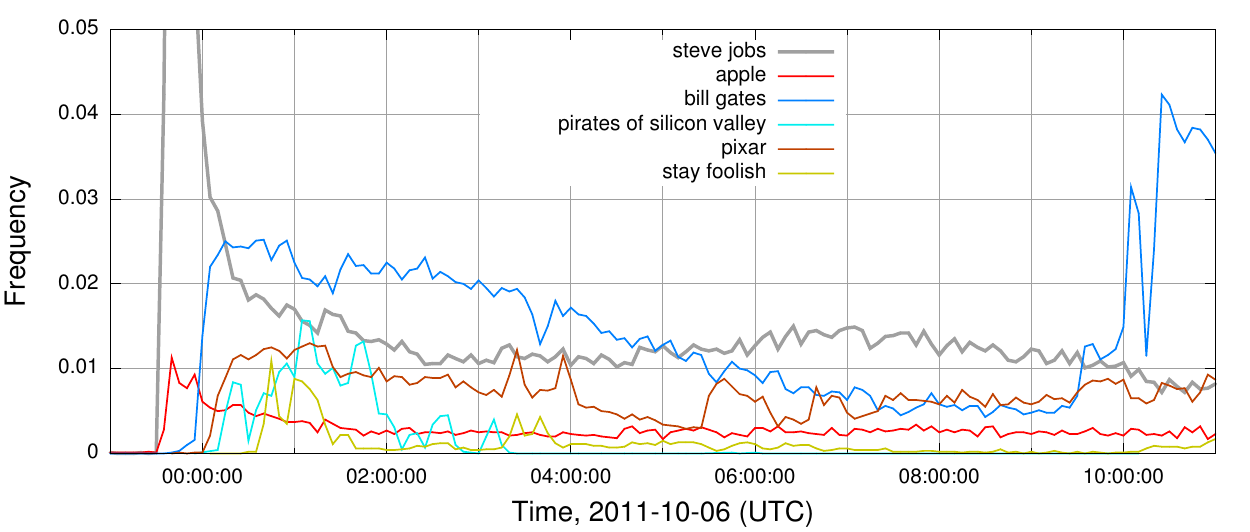}
\caption{Frequencies of queries related to Steve Jobs' death over a 12
  hour period in 5-minute intervals, normalized to the total number of
  queries in the interval.  At its peak, the query ``steve jobs''
  reached 15\% of the query stream; for readability of the
  other query frequencies, the scale is not stretched to include that
  point. (Reprinted from~\cite{Lin_Mishne_ICWSM2012})}
\label{figure:stevejobs-5-min}
\end{figure}

When the search assistance project began, one of our first tasks was
to better understand the temporal dynamics of user queries on Twitter
and to more precisely quantify the real-time requirements of the
related query suggestion application. These results were separately
published~\cite{Lin_Mishne_ICWSM2012},\footnote{\small
  http://engineering.twitter.com/2012/06/studying-rapidly-evolving-user.html}
but here we provide a summary.  Our study of how user interests
rapidly evolve focused on ``churn'', informally characterized as the
process by which terms and queries become prevalent and then ``drop
out of the limelight''. We refer the reader to the full paper, but
here we provide some highlights:

Examining all search queries from October 2011, we see that, on
average, about 17\% of the top 1000 query terms from one hour are no
longer in the top 1000 during the next hour. In other words, 17\% of
the top 1000 query terms ``churn over'' on an hourly basis.  Repeating
this at the granularity of days instead of hours, we find that
about 13\% of the top 1000 query terms from one day are no longer in
the top 1000 during the next day.

During major events, the frequency of queries increase dramatically. For
example, on October 5, immediately following the death of
Apple co-founder and CEO Steve Jobs, the query ``steve jobs'' spiked
from a negligible fraction of the query volume to 15\% of the query
stream. Figure~\ref{figure:stevejobs-5-min} (reprinted
from~\cite{Lin_Mishne_ICWSM2012}) shows the frequency of a few queries
related to Steve Jobs as a function of time. A short while
after the query ``steve jobs'' spiked, related queries such as
``apple'' and ``stay foolish'' (one of his mottos) spiked as well.

One of our conclusions from this study was that for real-time
applications on Twitter, we need to keep track of term statistics at a
fine granularity. A window of approximately five minutes seems to
be the sweet spot in being able to reflect large-scale changes; any
longer we'd be reacting too slowly, but any shorter it would be
difficult to accumulate sufficient counts for anything other than the
head of the vocabulary distribution. Although the death of Steve Jobs
is an extraordinary event, this episode is typical of how fast Twitter
reacts to significant breaking news. From this, we established a
target of returning relevant related query suggestions within
ten minutes after an event has occurred.

\subsection{Algorithm Overview}

At the core of Twitter's search assistance is a simple idea:\ if query
$A$ and query $B$ are seen in the same context, it provides evidence
that they are related. In many cases $A$ precedes $B$ in time:\ this
suggests that $B$ may be a query that is interesting to searchers who
found $A$ interesting.  Furthermore, if $A$ and $B$ are very similar,
as measured, for example, by edit distance, $B$ is likely a
spell-corrected version of $A$ (especially if $A$ returns far fewer
results than $B$ or no results at all).  Naturally, we accumulate
evidence across many different contexts before surfacing a suggestion
to the user.

This simple idea admits a large design space for instantiating the
actual algorithm. First, how do we define ``context''?  Presently, we
rely on two different types of context:\ a user's search session and
tweets themselves. User search sessions that span multiple queries
provide valuable signal---for example, the user might issue query $A$,
browse the results and notice something interesting that leads to
query $B$. This is exactly the type of connection that we want to
learn. Terms that appear together in tweets also provide valuable
evidence---this is closely related to the vast computational
linguistics literature on extracting collocations and other
semantically-related terms~\cite{Manning_Schutze_1999,Pearce02}.

After defining the context, the next question:\ how do we
quantitatively measure how often $A$ and $B$ appear together in the
same context? Once again, there is a large number of metrics to choose
from:\ conditional relative frequency, pointwise mutual information,
log-likelihood ratio, the $\chi^2$ statistic, just to name a few
popular ones.

Of course, we need to take into account the temporal aspects of the
evidence we have observed. There are several ways to accomplish this,
and our general approach is to ``decay'' observed counts over time,
which affects correlation statistics and gradually lessen the
importance of observed events as they age. However, even this simple
decay strategy leads to a wide range of choices for the decay
function:\ exponential, step-function, linear are obvious choices,
each with free parameters to tune (e.g., the $\alpha$ decay constant
for exponentials or the slope for linear decay).

Finally, we require a mechanism to combine all the evidence from each
individual relevance signal (i.e., a ranking algorithm). The simplest
workable strategy is a linear combination, with either hand-tuned or
machine-learned weights, but here is an opportunity to leverage
learning-to-rank techniques~\cite{LiHang_2011} such as
gradient-boosted regression trees and ensemble
methods~\cite{Ganjisaffar_etal_SIGIR2011}. Note that in reality the
production system runs multiple algorithms, either as part of A/B
testing experiments or as part of ensembles whose results are then
combined.

The purpose of this description is to illustrate the types of signals
and features that are exploited by the search assistance service
without describing the actual algorithm (which we hope to detail in a
future paper). However, we believe this outline provides the
reader with sufficient context to understand the remainder of the
paper and appreciate the architectural challenges involved in this
problem. At an abstract level, the relevance signals can be thought of
as the problem of computing functions over the query space crossed
with itself (i.e., all possible $A$'s crossed with all possible $B$'s;
cf.\ the ``All-Pairs'' problem~\cite{Moretti_etal_2008}). In theory,
the query space is the power set of the vocabulary space (ignoring
queries that return zero results), but in practice queries are short,
and we only consider $n$-grams up to $n=3$. However, even with this
simplification the event space is quite large and cannot be fully
materialized in memory. We return to this issue later.

Our general approach is closest to the session-based technique
described by Jones et al.~\cite{JonesRosie_etal_WWW2006}, but only
begins to scratch the surface in terms of algorithms that can be
brought to bear in tackling the real-time related query suggestion
problem. Although not currently implemented yet, we have given some
thought to how more sophisticated algorithms, such as random walks on
the query-clickthrough
graph~\cite{Mei_etal_CIKM2008,CaoHuanhuan_etal_SIGKDD2008}, can be
adapted to the real-time context, but leave these interesting
enhancements for future work.

\section{Take One:\ Hadoop Solution}

When the search assistance project began, the most obvious solution
was to take advantage of the existing analytics platform for data
processing. Over the past several years, Twitter has built a robust,
production petabyte-scale analytics platform, primarily based on
Hadoop, but also incorporating other components such as Pig, HBase,
ZooKeeper, and Vertica. The first (complete) version of search
assistance was built using this platform, but was later
replaced. Before discussing the reasons for this, we provide a brief
overview our Hadoop platform here, and refer the reader to
previously-published papers for more
details~\cite{Lin_etal_MAPREDUCE2011,Lin_Kolcz_SIGMOD2012,Lee_etal_VLDB2012}.

A large Hadoop cluster lies at the core of the analytics
infrastructure that serves the entire company. Data is written to the
Hadoop Distributed File System (HDFS) via a number of real-time and
batch processes, in a variety of formats. These data can be bulk
exports from databases, application logs, and many other sources.
When the contents of a
record are well-defined, they are serialized using either Protocol
Buffers\footnote{{\small http://code.google.com/p/protobuf/}} or
Thrift,\footnote{{\small http://thrift.apache.org/}} and typically
LZO-compressed. We have written an open-source system called Elephant
Bird,\footnote{{\small http://github.com/kevinweil/elephant-bird}}
that hooks into the serialization frameworks to automatically generate
code for reading, writing, and manipulating arbitrary Protocol Buffer
and Thrift messages.

Instead of directly writing Hadoop code in Java, analytics at Twitter
is performed mostly using Pig, a high-level dataflow language that
compiles into physical plans that are executed on
Hadoop~\cite{Olston_etal_SIGMOD2008,Gates_etal_VLDB2009}. Pig provides
concise primitives for expressing common operations such as
projection, selection, group, join, etc. This conciseness comes at low
cost:\ Pig scripts approach the performance of programs directly
written in Hadoop Java. Yet, the full expressiveness of Java is
retained in the ability to call arbitrary user-defined functions
(UDFs).

Production Pig analytics jobs are coordinated by our workflow manager
called Oink, which schedules recurring jobs at fixed intervals (e.g.,
hourly, daily). Oink handles dataflow dependencies between jobs; for
example, if job \textit{B} requires data generated by job \textit{A},
then Oink will schedule \textit{A}, verify that \textit{A} has
successfully completed, and then schedule job \textit{B} (all while
making a best-effort attempt to respect periodicity
constraints). Finally, Oink preserves execution traces for audit
purposes:\ when a job began, how long it lasted, whether it completed
successfully, etc. Each day, Oink schedules hundreds of Pig scripts,
which translate into thousands of Hadoop jobs.

The first version of search assistance was written in Pig, with custom
Java UDFs for computations that could not be directly expressed with
Pig primitives. A Pig script that aggregates user search sessions,
computes term and co-occurrence statistics, and ranks related queries
and spelling suggestions would run on
our Hadoop stack; a lightweight frontend periodically loaded the
output and served the results for incoming requests.

The system worked reasonably in terms of output quality and allowed us
to experiment and discover useful signals, but the latency was
unacceptable. Related query suggestions were not available
until several hours after the collection of the data those suggestions
were based on. Initially, we were somewhat surprised by this lag and
spent some effort to understand the issues involved. Next, we detail
the two primary bottlenecks and why they existed.

\subsection{Bottleneck One:\ Log Import}
\label{section:1:scribe}

The first bottleneck involved the data import pipeline---moving log
data from tens of thousands of production hosts onto HDFS. In
particular, search assistance made use of ``client event''
logs, which capture records of users interactions across the various
Twitter clients (e.g., the twitter.com site, iPhone and Android apps,
etc.). These logs, on the order of a terabyte a day (compressed) as of
summer 2012, capture everything from site navigation to page
impressions, and of course, include the query contexts for search
assistance. For additional details on the client event logging
infrastructure, we refer the reader to a recent
paper~\cite{Lee_etal_VLDB2012}.

\begin{figure}[t]
\centering\includegraphics[width=0.95\linewidth]{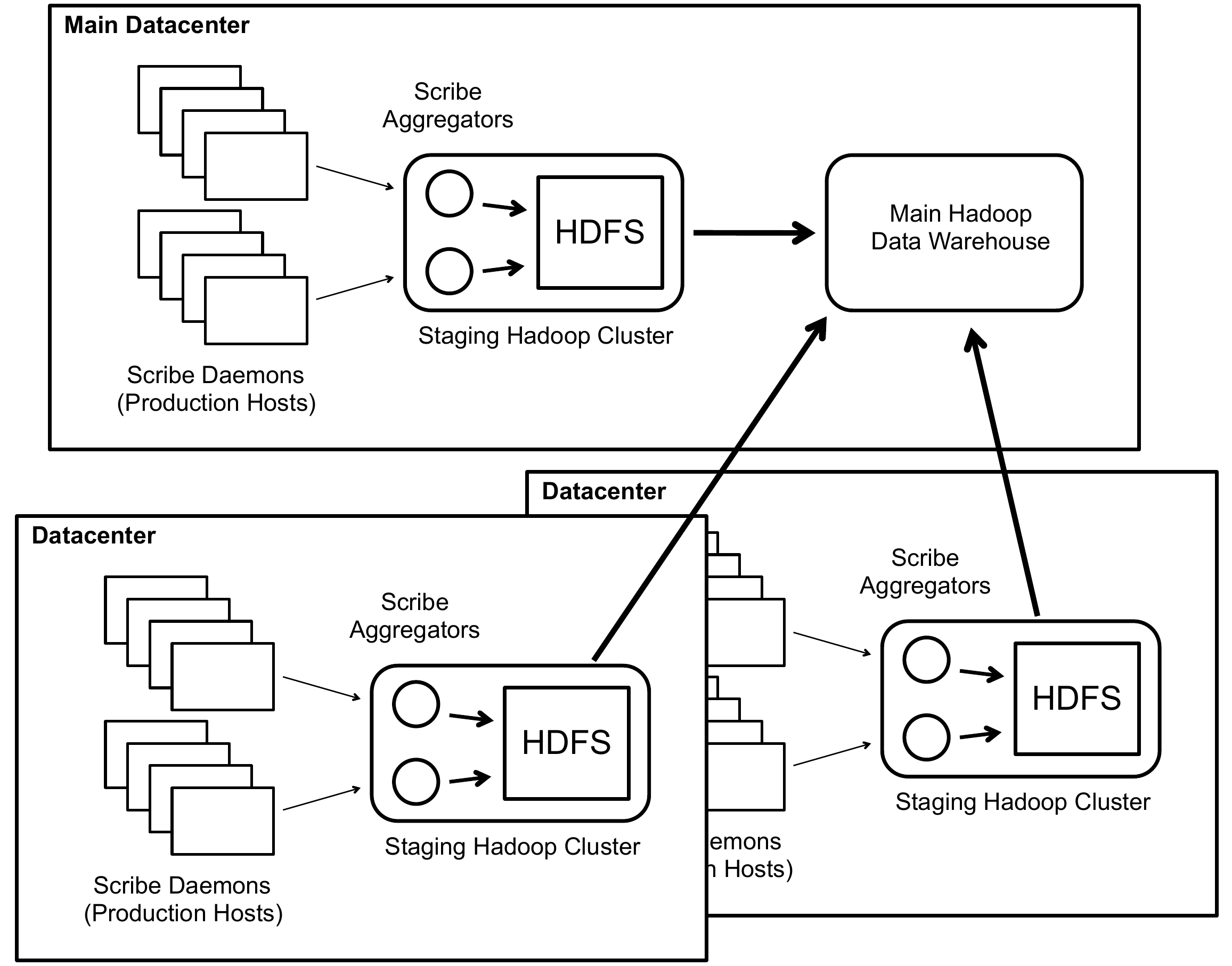}
\caption{Illustration of Twitter's Scribe infrastructure. Scribe
  daemons on production hosts send log messages to Scribe aggregators,
  which deposit aggregated log data onto per-datacenter staging Hadoop
  clusters. Periodic processes then copy data from these staging
  clusters into our main Hadoop data warehouse.}
\label{figure:scribe}
\end{figure}

For gathering log data, Twitter uses Scribe, a system for aggregating
high volumes of streaming log data in a robust, fault-tolerant,
distributed manner. It was originally developed and later open sourced
by Facebook. Although it has since been augmented by other systems,
Scribe remains an integral part of Facebook's logging
infrastructure. Twitter's Scribe infrastructure is illustrated in
Figure~\ref{figure:scribe}, and is similar to the design presented
in~\cite{Thusoo_etal_SIGMOD2010}. A Scribe daemon runs on every
production host and is responsible for sending local log data across
the network to a cluster of dedicated aggregators in the same
datacenter. Each log entry consists of two strings, a category and a
message. The category is associated with configuration metadata that
determine, among other things, where the data is written.

The aggregators in each datacenter are co-located with a staging
Hadoop cluster. Their task is to merge per-category streams from all
the server daemons and write the merged results to HDFS (of the
staging Hadoop cluster), compressing data on the fly.  Another process
is responsible for moving these logs from the per-datacenter staging
clusters into the main Hadoop data warehouse. It applies certain
sanity checks and transformations, such as merging many small files
into a few big ones and building any necessary indexes. Lastly, it
ensures that by the time logs are made available in the main data
warehouse, all datacenters that produce a given log category have
transferred their logs. Once all of this is done, the log mover
atomically slides an hour's worth of logs into the main data
warehouse. At the end of the log mover pipeline, logs arrive in the
main data warehouse and are deposited in per-category, per-hour
directories (e.g., {\tt \small /logs/category/YYYY/MM/DD/HH/}). Within
each directory, log messages are bundled in a small number of large
files. From here, our Oink workflow manager fires off a cascade of Pig
jobs that compute the related query suggestions.

Unfortunately, there is a substantial delay from when the logs are
generated to when they are available in the main data
warehouse. Typically, we observe lag on the order of a couple of
hours, although delays of up to six hours are not uncommon. This
clearly does not meet the real-time demands of our application.

It is important to note that our Scribe architecture adopts standard
best practices in the industry, and there are good reasons for each
aspect of the design. The hierarchical aggregation scheme is necessary
because HDFS cannot handle large numbers of small files---otherwise, a
simpler design is to have production hosts directly write logs into
HDFS. The aggregators allow log data from many Scribe daemons to be
``rolled up'' into a smaller number of large files---this also
provides a hook for ETL operations such as compression, data cleaning,
building indexes, etc.

Data import is also bounded by the slowest task to complete, because
the process was designed to appear atomic to downstream consumers. For
example, when the workflow scheduler Oink observes that a
newly-created hourly log directory appears, it assumes that all logs
are present. This assumption simplifies the design of Oink
in not having to deal with partially transferred data. It is not uncommon
for some aggregators to lag a bit behind, perhaps due to an
idiosyncratic distribution of Scribe daemons that are connected to
it. Furthermore, the log mover operates across
geographically-distributed datacenters, and therefore is subjected to
the uncertainties of copying large amounts of data over a wide-area network.

There are possible ways to reduce the latency in the log import
pipeline within the existing Scribe architecture. We could, for
example, implement sub-hour incremental importing. This would come at
the cost of additional complexity to data consumers since we'd need a
signaling mechanism to notify that all data for a particular hour has
arrived. This is not impossible, but would require substantially
re-engineering the analytics stack. Incremental importing, however,
might exacerbate the small files problem in HDFS---we still
need to accumulate log data over some interval to avoid a
proliferation of small files. In the best case, we could probably
achieve latencies in the tens of minutes from when the logs are
generated to when they are available on HDFS for processing. This latency
still remains too high for our application.

We are aware that since the development of Scribe there have been
advances that tackle the issue of real-time log processing, for
example, Facebook's ptail and Puma
combination~\cite{Borthakur_etal_SIGMOD2011} and LinkedIn's
Kafka~\cite{Kreps_etal_2011,Goodhope_etal_2012}. We return to discuss
this in more detail in Section~\ref{section:future}.

\subsection{Bottleneck Two:\ Hadoop}

The second bottleneck in our initial implementation on the Hadoop
analytics platform had to do with the latencies associated with
MapReduce jobs themselves. There were two issues, discussed in
detail below:

The first issue involved contention on the Hadoop cluster, which is a
shared resource across the company. On a typical day, it runs tens of
thousands of {\it ad hoc} and production jobs from dozens of
teams around the company. We use the Fair\-Scheduler, which does have a notion of task
pools with different priorities. However, this is not the best
mechanism for our purposes, since we don't care about resource
allocation as much as having predictable bounds on end-to-end job
completion times. Perhaps as a testimony to the success of ``big
data'' analytics, cluster usage outpaces the growth of physical
cluster resources.

The second issue involved the speed of MapReduce jobs and the
complexity of the search assistance algorithm itself. An initial
prototype in Pig translated into roughly a dozen MapReduce jobs and
took around 15--20 minutes to process one hour of log data (without
resource contention). Complexity of the algorithm aside (some of which
was unavoidable to generate high quality results), there were several
contributing factors to the slow speed:\ Hadoop was simply not
designed for jobs that are latency sensitive. On a large cluster, it
can take tens of seconds for a large job to start up, regardless of
the amount of data it processes.

Another issue we observed was the sensitivity of job completion times
to stragglers. Many aspects of natural language, for example, the
distribution of vocabulary terms, follow Zipfian distributions, for
which a simple hash partitioning scheme creates chunks of ``work''
that vary greatly in size. For many jobs, the mean running time for
map or reduce tasks was substantially shorter than the maximum (which
bounds the job completion time). In a complex Pig script that chains
together a dozen or more Hadoop jobs, the end-to-end critical path is
the slowest task in each MapReduce cycle. The issue of stragglers has
been studied by
researchers~\cite{AnanthanarayananGanesh_etal_OSDI2010,Kwon_etal_SIGMOD2012},
but even in Pig there are a number of ways to address the issue with
careful software engineering---for example, properly setting the {\tt
  \small parallel} factor (albeit, a brute force solution), using
certain types of joins when appropriate, etc.

However, even with careful software engineering, the best case
scenario for computing query suggestions is on the order of ten
minutes (without resource contention), due to the need to compute many
features and the amount of data involved. Coupling this with the delay
from log import, we arrive at, optimistically, an end-to-end latency
of a few tens of minutes. Based on the analysis presented in
Section~\ref{section:bg:churn}, this does not seem quick enough---by
the time we begin to make relevant related query suggestions, the
breaking news event might have already passed us by.

\subsection{Revisiting the Decision to Use Hadoop}

While the issues with the Hadoop implementation seem obvious with the
benefit of hindsight, several factors made the Pig implementation a
natural starting point.

First, because Twitter already had a mature Hadoop-based analytics
platform, implementing the search assistance algorithms as Pig scripts
required no additional infrastructure, and a working prototype was
built within a short time. Various aspects of building production Pig
workflows using Oink, such as scheduling, resource management, error
handling, notifications, etc., are well established. Therefore, it was
easy to immediately get started and rapidly iterate.

Second, and related to the first, the search team had already written
a large number of Pig scripts that analyzed search logs. These range
from relatively simple aggregation jobs that fed frontend dashboards
(fully productionized, running daily) to sophisticated {\it ad hoc}
analyses that were designed to answer some specific question. There
was a large body of code we could borrow from to serve as the basis of
the initial search assistance prototype.

Third, when we started the project we did not yet have the in-depth
understanding of query churn on Twitter that we described in
Section~\ref{section:bg:churn}.  One of the benefits of the initial
system, in addition to code and data that could be reused later,
was insight into the rapid changes in the query stream, which forced
us to focus on a more real-time solution.

\section{Take Two:\ Deployed Solution}

Although we eventually replaced the Hadoop-based architecture
described in the previous section due to its inability to meet the
latency requirements of the search assistance application, in no way
did we consider it a ``failure''. Quite the contrary, developing on
the Hadoop-based analytics stack allowed us to experiment on a large
amount of retrospective data and to conveniently explore the algorithm
design space. Although {\it architecturally}, the deployed solution
was completely redesigned, many of the {\it algorithms} and some of
the code (e.g., inside Pig UDFs)\ remained unchanged.

One advantage of the Hadoop-based architecture was its generality,
since it had access to logs that captured a wide range of user
interactions---not only searches and tweet activity, but also
impressions, clicks, etc. This in theory allowed us to deploy very
sophisticated algorithms, including those that operate on
clickthrough graphs and those that take into account Twitter idioms
such as retweets, replies, and favorites. However, we discovered that
using two sources of context---search sessions and tweets---were
sufficient to provide good results, at least for an initial
implementation. Thus, our deployed solution amounted to a custom
in-memory processing engine that focused on these two sources of data,
augmented by offline processing components.

The remainder of this section describes the architecture of our
deployed system, which then sets up our discussion of future work in
Section~\ref{section:future}. The limitations of what we have built
gives us some idea of where the gap is between processing ``big'' and
``fast'' data.

\subsection{Overall Search Architecture}

\begin{figure}[t]
\centering\includegraphics[width=0.65\linewidth]{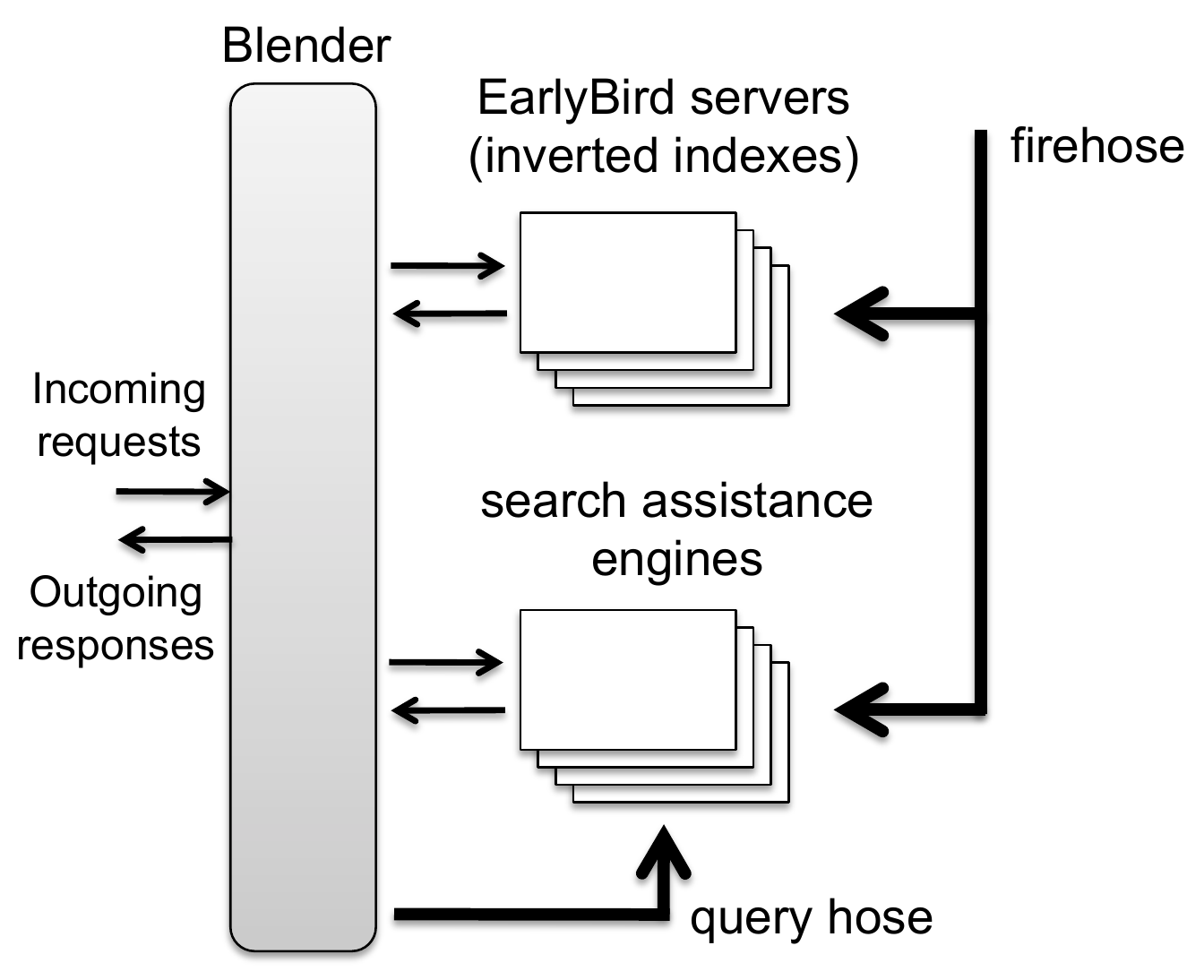}
\caption{Twitter's overall search architecture, showing the Blender,
  which brokers access to all search services, including the EarlyBird
  servers and the search assistance engines. Since the Blender
  receives all search queries from Twitter clients, it is able to
  provide a ``query hose'' to the search assistance engines.}
\label{figure:search}
\end{figure}

The relevant parts of the overall Twitter search architecture is shown
in Figure~\ref{figure:search}. A frontend called the search
``Blender'' brokers all requests (for example, from the twitter.com
web client) to Twitter's family of search services (searching for
tweets, searching for user accounts, search assistance, etc.).
EarlyBird~\cite{Busch_etal_ICDE2012} is the name of our inverted
indexing engine. A fleet of these servers ingests tweets from the
``firehose''---a streaming API providing access to all tweets as they
are published---to update in-memory indexes.

Because of this architecture, the Blender has a complete record of
users' search sessions---there is no need for client event Scribe logs
(as detailed in Section~\ref{section:1:scribe}). The Blender makes
these queries (and associated session data) available internally as a
``query hose'' service, akin to the firehose. Note, however, data
available to the Blender is relatively limited; for example, it
doesn't have access to clickthrough data. This is a limitation in the
future when we wish to augment the search assistance algorithm to take
advantage of more relevance signals.

\subsection{The Search Assistance Engine}

\begin{figure}[t]
\centering\includegraphics[width=0.95\linewidth]{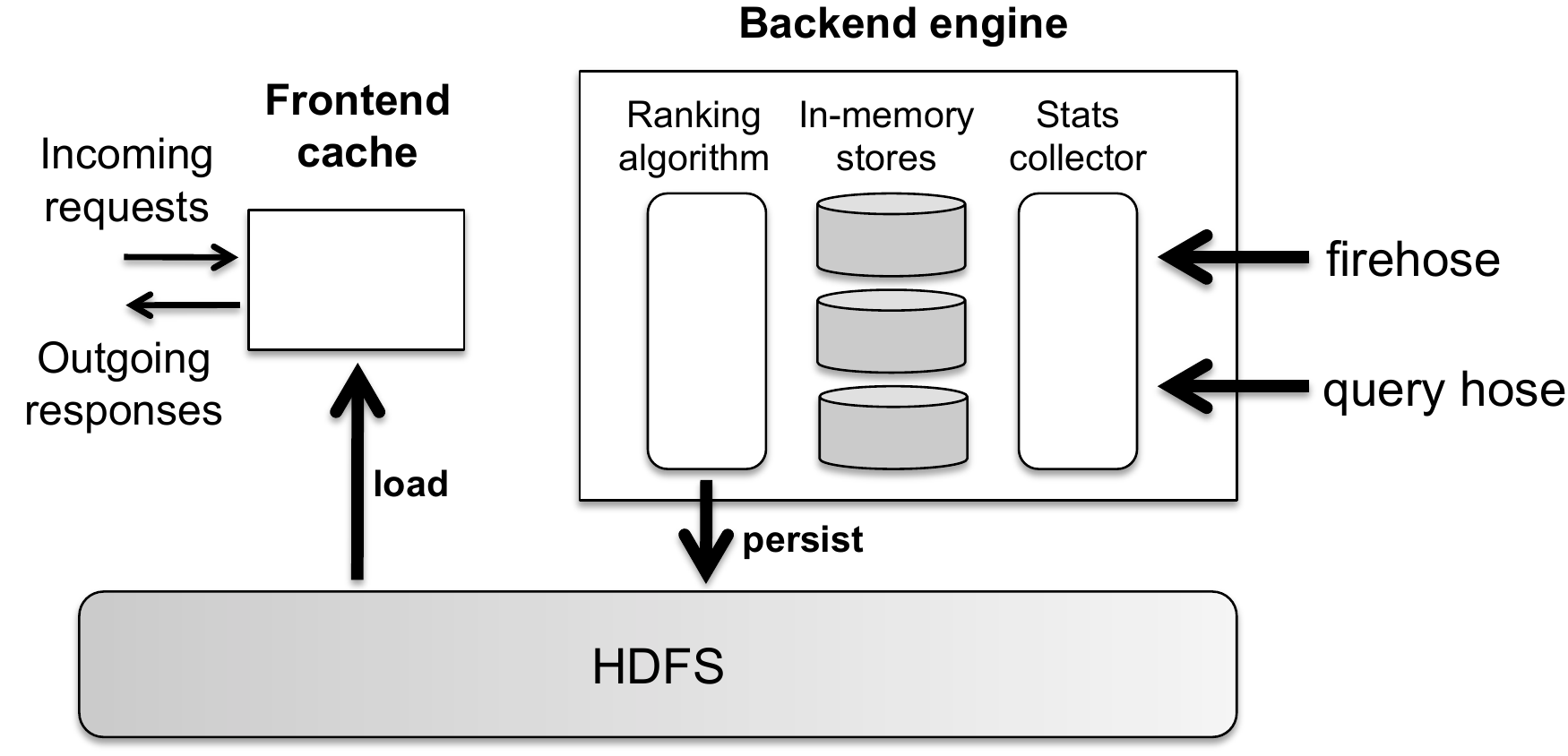}
\caption{Architecture of the search assistance engine, consisting of a
  lightweight frontend serving requests from an in-memory cache and a
  backend that consumes the firehose and query hose to compute related
  query suggestions and spelling corrections.}
\label{figure:engine}
\end{figure}

Twitter search assistance is provided by a custom, in-memory
processing engine that consumes two sources of input:\ the tweet
firehose and the Blender query hose (as described above). The design
is shown in Figure~\ref{figure:engine} and comprises two decoupled
components:\ lightweight in-memory caches, which periodically read
fresh results from HDFS, serve as the frontend nodes, while actual
computations are performed on backend nodes.  Each of the frontend
caches is implemented as a Thrift service, and together they form a
single replicated, fault-tolerant service endpoint that can be
arbitrarily scaled out to handle increased query load. Request routing
to the replicas is handled by a Twitter abstraction called a
Server\-Set, which provides client-side load-balanced access to a
replicated service, coordinated by
ZooKeeper~\cite{HuntPatrick_etal_2010} for automatic resource
discovery and robust failover. Details about the Server\-Set
abstraction were presented in a previous
paper~\cite{Leibert_etal_SoCC2011}.

The backend processing engine is replicated for fault tolerance, but
not sharded (i.e., each instance independently holds the entire state
of the application). Every five minutes, computed results are
persisted to HDFS---the instances perform leader election using
ZooKeeper, and the winner proceeds to write its results. Every minute,
the frontend caches poll a known HDFS location for updated results,
thus ensuring freshness of query suggestions and spelling corrections.

We adopted this decoupled frontend and backend design for a few
reasons:\ first, due to different scalability requirements. The
frontend caches need to scale out with increased query load, whereas
the backends face no such pressure---we simply need to guarantee
sufficient fault tolerance so that {\it some} instance persists
updated results every five minutes. Persisting data to HDFS has many
advantages:\ upon a cold restart, the frontend caches can serve the
most-recently written results immediately without waiting for the
backend. Persisted results can also be analyzed retrospectively with
Pig to better understand how the service is being used---since the
data are on HDFS, it is easy to join them with log data for behavior
analysis and user data to slice-and-dice by demographic
characteristics.

The search assistance frontend implementation is fairly
straightforward so we leave out further details. Each backend instance
is a multi-threaded application that consists of three major
components:\ the stats collector, which reads the firehose and query hose,
in-memory stores, which hold the most up-to-date statistics, and
rankers, which periodically execute one or more ranking algorithm by
consulting the in-memory stores for the raw features.

There are three separate in-memory stores to keep track of relevant
statistics:

\smallskip \noindent The {\it sessions store}, which keeps track of
(anonymized) user sessions observed in the query hose, and for each,
the history of the queries issued in a linked list. Note that we only
keep a limited history of session data, so sessions older than a
threshold are discarded; effectively, the sessions are sliding windows
holding the users' most recent search activity. Separately, we keep
track of metadata about each session:\ the unique queries contained in
each session, the unique query cooccurrence pairs, and so on.

\smallskip \noindent The {\it query statistics store}, which retains
up-to-date statistics about individual queries. These include the
count of the number of sessions they've been observed in as well as a
weighted count based on a custom scoring function. Scoring is used to
capture some Twitter-specific aspects of search:\ for example, queries
may originate from different sources, such as typing in the
search box, clicking a hashtag, or clicking a related query. The
association strength between consecutive queries depends on
their sources:\ intuitively, two hashtag clicks are not as strongly
indicative as consecutive typed-in queries, and this is reflected
in the incremental weight added to individual queries
when a query instance is observed. In this store we also periodically
decay weights to reflect decreasing importance over time, in the
absence of additional statistical evidence from the query hose.
Finally, we keep additional metadata about the query such as its
detected language for the purpose of serving different results in
different international markets.

\smallskip \noindent The {\it query cooccurrence statistics store},
which is similar to the query statistics store, except that it holds
data about {\it pairs} of cooccurring queries (stored in a sparse data
structure). We apply a similar weighting and decay scheme as above. In
addition, for each query, we store all unique queries that follow it
in at least one session, and all unique queries that precede it in at
least one session. Note that we use a single data structure to keep
track of cooccurrences in both search sessions and tweets. There are
naturally many more cooccurring terms in tweets, but we do not keep
track of pairs that are not observed in queries, which significantly
reduces the event space.

\subsection{Data Flow}

In more detail, the following takes place in a search assistance
backend node.

\smallskip \noindent {\em The query path}:\ as a query from a given
user is delivered through the query hose, the following actions are
taken:
\begin{list}{\labelitemi}{\leftmargin=1em}
\setlength{\itemsep}{-2pt}

\item Query statistics are updated in the query statistics store:\ raw
  counts and scored weights based on interaction type (e.g., click
  on hashtag, typed-in).

\item The query is added to the sessions store. A new session is
  created if necessary. If needed, old queries are removed from the
  session to preserve the sliding window size.

\item For each previous query in the session, a query cooccurrence is
  formed with the new query. Statistics are updated in the query
  cooccurrence statistics store accordingly.
\end{list}
\noindent Note that once session statistics are available, new
queries are subject to rate-limiting and other checks.

\smallskip \noindent {\em The tweet path}:\ As a tweet is delivered
through the firehose, all {\it n}-grams from it are processed to
determine whether they are query-like or not (i.e., whether they are
observed often enough as standalone queries). All {\it n}-grams that
match queries are processed in a similar way to the query path above,
except that the ``session'' is the tweet itself.

\smallskip \noindent {\em Decay/Prune cycles}:\ Periodically, all
weights (queries and query cooccurrences) are decayed; queries or
cooccurrences with scores falling under predefined thresholds are
removed to control the overall memory footprint of the
service. Similarly, user sessions with no recent activity are pruned.

\smallskip \noindent {\em Ranking cycles}:\ In a separate periodic
process, a particular ranker (consisting of an algorithm and the
parameters for its execution) is triggered. The ranker traverses the
entire query statistics store and generates suggestions for each query
based on the various accumulated statistics; top results are then
persisted to HDFS.

\subsection{Scalability}

There are two scalability bottlenecks in our design. The first is the
fact that each instance of the backend processing engine must consume
the entire firehose and query hose. Since there is no partitioning of
the data streams, a single server must keep up with the incoming
data. The stats collector is multi-threaded, with threads in two
separate thread pools pulling from the firehose and query hose. In our
benchmarks, CPU is not a limiting resource, and we appear to have
plenty of headroom for the foreseeable future.

The other scalability bottleneck is memory for retaining the various
statistics, particularly since the event space for cooccurring queries
is quite large without any pruning. This is not merely a performance
issue, as it can have substantial impact on the quality and coverage
of the results. The coverage versus memory footprint tradeoff is
fairly obvious:\ we can reduce memory consumption by only keeping
track of frequently-occurring query terms (above a threshold), but at
the cost of coverage, i.e., for how many queries we can generate
meaningful suggestions. Another approach to reducing memory footprint
is to store less session history and more aggressively decay
weights. However, these are exactly the decisions that impact result
relevance. Ideally, we should be able to isolate algorithmic parameter
setting from performance considerations, but in reality both are at
least partially intertwined.

\subsection{Background Models}

The search assistance engine described above tracks recent queries in
real-time, but has limited temporal coverage. Statistics belonging to
queries that are more than a day old have sufficiently decayed to a
point where their impact on the final results is negligible, or have
been completely pruned out of the stores.  To boost query coverage
(the number of queries for which we generate suggestions or spelling
corrections), we have a couple of additional mechanisms.

The first involves running the same search assistance
backend, except over data spanning much longer periods of time (on the
order of several months), but with different parameter settings
(decay, pruning, etc.). These processes run periodically (currently,
every six hours) and provide a ``background model'' to capture
slower-moving trends and suggestions that are persistent over time.

As another useful feature for spelling correction, we perform a
pairwise edit distance variant calculation between all queries
observed within a long span of time (the variant accounts for some
spelling-specific issues, such as mistakes being more frequently
observed in internal characters of a word rather than at the beginning
or the end, as well as accounting for Twitter specifics such
as\ @mentions and hashtags).  This captures misspelling such as
``justin biber'', which is common and persistent. For this, we run a
Pig job.

The results of these less-frequent jobs are also deposited on
HDFS:\ the frontend caches load them and perform interpolation with
the real-time results to serve the final output.

\section{Future Directions}
\label{section:future}

In the process of twice implementing the search assistance service, we
gained experience working with ``fast data'' and the limitations of a
Hadoop-based analytics stack. The custom deployed solution works well
but is inflexible. We do not believe that the requirements of related
query suggestion are unique, but rather represent a gap between
platforms for processing ``big data'' and ``fast data'' in general.
In this section, we attempt to better articulate an important future
direction in data management and discuss current work in this
direction.

\subsection{The General Problem}

At an abstract level, we desire a {\it general} and {\it unified} data
processing framework that can execute complex queries involving
arbitrary user-specified computations, at varying levels of temporal
granularity, with varying latency requirements. In the case of related
query suggestion, we need to compute functions over the space of user
search queries crossed with itself, since we're accumulating evidence
on {\it pairs} of queries. The results of these computations are then
combined by the ranking algorithm, but this part is relatively
straightforward compared to computing, storing, and updating the raw
statistics. The challenge lies in the fact that we need statistics
across temporal granularities that differ by several orders of
magnitude. We need evidence at the minute-by-minute level to track
fast moving, breaking events, as well as evidence accumulated across
days, weeks, or even months for slower moving and tail queries.

Our deployed solution is neither general nor does it represent a
unified processing framework. We were able to build a custom in-memory
processing engine for search assistance because we learned from the
Hadoop implementation that two signals (tweets and search
sessions)\ were sufficient to generate good results---the fact
that the search Blender had access to the query stream made it easier
to feed session data directly to the search assistance
backend. However, we are unable to exploit (without additional custom
workarounds) the far richer sources of signal in the full client event
logs---clicks, impressions, etc.

Furthermore, the deployed system remains a patchwork of different
processing paradigms:\ the search assistance engine running on
real-time data, the same engine running on larger amounts of
retrospective data, and Pig jobs handling the long tail of query
misspelling. This situation is far from ideal, since it results in
code duplication and increased complexity from coordinating multiple
processes.

The need to compute statistics across very different temporal
granularities with different latency tolerances creates additional
processing constraints. Real-time processing generally implies holding
all data in memory. In many cases this is not possible, thus
necessitating approximations or pruning to avoid out-of-memory
errors. On the other hand, batch computations on Hadoop generally do
not have this limitation since intermediate data are materialized to
disk, and because there are far less stringent latency requirements
we can afford to compute statistics for the entire long
tail. Currently, we must manage these constraints by hand---for
example, hand tuning pruning and decay parameters depending on how
much data we are processing. Once again, it would be desirable for a
data processing framework to ``figure out'' these issues and adapt a
query plan accordingly.

Although the challenges we sketched out are couched in the context of
related query suggestion, these issues are certainly not unique to
us---for example, the literature discusses real-time computation of
clickthrough rates (CTR), particularly in the context of online
advertising~\cite{Neumeyer_etal_2010,Chandramouli_etal_ICDE2012}. The
nature of the marketplace demands that ad placement algorithms have
access to the most recent statistics. However, there is often a need
to perform analytics over longer periods of time (e.g., across days or
weeks) to uncover underlying trends. In most setups, some sort of
online processing engine is used for the real-time case, and a batch
analytics platform for the latter case. It would be desirable to have
a single {\it unified} data processing platform that ``does it all''.

\subsection{Pieces of the Solution}

One important future direction in data management is bridging the gap
between platforms for ``big data'' and ``fast data''. We believe that
pieces of the solution already exist, but to our knowledge there
hasn't been anything published that integrates everything into a
unified data processing framework. We discuss some relevant work:

Large-scale publish-subscribe systems such as Hedwig\footnote{\small
  http://wiki.apache.org/hadoop/HedWig} and
Kafka~\cite{Kreps_etal_2011,Goodhope_etal_2012} present nice solutions
to the problem of moving large amounts of data around in a robust and
scalable manner. According to LinkedIn~\cite{Goodhope_etal_2012},
Kafka handles more than 10 billion message writes each day with a
sustained peak of over 172,000 messages per second. For real-time
processing, this seems like a superior solution to Scribe. However,
Kafka alone is not sufficient, as it lacks a processing engine and the
ability to persist data over long spans (but in fairness, the system
was not designed for those two tasks). In LinkedIn's architecture,
there is a process that consumes Kafka messages and persists them to HDFS
at ten-minute intervals (presumably, because of the small file
problem). Even with this architecture we would be unable to meet our
freshness requirements. For search assistance the target latency is ten
minutes {\it end-to-end} (including data processing and candidate
ranking), which would still preclude a Kafka/Hadoop solution.

Interestingly, Facebook adopts a completely different architecture
with a combination of ptail and
Puma~\cite{Borthakur_etal_SIGMOD2011}. On top of a Scribe
infrastructure, Facebook has implemented ptail, which is like the Unix
``tail'' command, except for HDFS data. A process runs ptail,
consuming the end of logs as they are written to HDFS, and pipes it to
Puma, which is their in-memory aggregation engine. Aggregates are
``flushed'' periodically to HBase, which is the system of record for
real-time results. Since this design has been in production at
Facebook, we assume that it scales in practice, although it is unclear
whether the solution is a clever hack or a general design that can be
elevated to the status of ``best practice''.

Stream-oriented databases have a long
history~\cite{Carney_etal_VLDB2002,Gehrke_2003,Gedik_etal_SIGMOD2008,KrishnamurthyS_etal_SIGMOD2010}. Typically,
users issue standing queries in a variant of SQL with temporal
extensions and results are returned via some sort of callback. One
advantage of these systems is that they build on widespread
familiarity of SQL by developers and data scientists. In addition,
most systems already have built-in primitives representing various
temporal constructs such as sliding windows, which makes a large class
of queries very easy to write (e.g., counting clicks and clickthrough
frequencies). In a similar vein, stream processing engines have
received renewed interest in the open source
community:\ S4~\cite{Neumeyer_etal_2010} and Storm\footnote{\small
  http://storm-project.net/} are two examples. However, we
see a few issues:\ It is unclear to what extent these systems address
the data persistence problem. For example, Storm and S4 do not provide
a built-in solution, other than having one of their processing
elements write to HDFS---but this begs the question of whether HDFS
should be the {\it source} as in the Facebook design or {\it sink} as
in the LinkedIn design. In general, stream-oriented databases
primarily operate in memory and were not designed to persistent large
amounts of data (if at all)---likely not the terabytes-of-data-per-day
scale that is common in popular web services.

The other issue with stream processing engines is that they are, for
the most part, not designed for queries with large temporal
spans. Handling a CTR calculation over a 30 second interval is surely
doable, but it is unclear whether they were designed for answering
similar types of queries over one week's worth of log data
(potentially tens of terabytes or more). In other words, although
stream processing engines excel at the real-time processing aspects,
it is not clear if they can handle more traditional complex {\it ad
  hoc} queries at a massive scale that is the bread and butter of
Hadoop-based stacks today. The recent work of Chandramouli et
al.~\cite{Chandramouli_etal_ICDE2012} in {\it embedding} a stream
processing engine inside a batch analytics framework appears to be a
step in the right direction.

Another interesting architecture that tries to address incremental online
computations at scale is Google's
Percolator~\cite{PengDaniel_Dabek_etal_OSDI2010}, which can be
summarized as database triggers for
Bigtable~\cite{ChangFay_etal_OSDI2006}. One application of Percolator
is incremental web indexing, which has elements of both velocity and
volume. We see, however, two potential issues for a Percolator-type
architecture as a general model of online data processing. First, it
assumes a Bigtable-like data model, and although such a data model is
fairly general, it is not appropriate for all cases. Second, by the
authors' own account, Percolator uses approximately 30 times more CPU
per transaction than a commercial DBMS on the TPC-E benchmark, which
seems costly to scale out, even with cheap commodity
servers. According to experiments reported in the paper, the system
achieves reads and writes in the tens of thousands per second range,
on a fairly large cluster:\ this falls short of the hundreds of
thousand of messages per second range needed for log processing at
scale (see Kafka performance statistics above). In fairness, we're
comparing apples to oranges, since Percolator supports multi-row
transactions, but such a consistency model is perhaps overkill for the
types of applications we're focused on. Although interesting,
Percolator occupies a different point in the design space.

Most recently, Lam et al.~\cite{LamWang_etal_VLDB2012} proposed
MapUpdate, an attempt to generalize MapReduce to
streams. Since streams may never end, ``updaters'' use storage called
{\it slates} to summarize the data they have seen so far, serving as
``memories'' of updaters, distributed across multiple machines and
persisted in a key--value store for later processing. The Muppet
implementation of MapUpdate focuses on how to efficiently execute
arbitrary code, but does not presently handle dynamic load
partitioning (except in event of machine failure). It also lacks a
higher-level query language for concisely expressing common
computations. While interesting and definitely a step in the right
direction, it is unclear if MapUpdate adequately covers all the use
cases we are interested in.


We envision a data processing framework that combines elements of a
stream processing engine to handle real-time computations and a
Hadoop-based batch analytics platform to perform ``roll ups'' and
handle large-scale analytical queries over long timespans. Internally,
we have been experimenting with various elements of the technologies
discussed above and have a few working prototypes that incrementally
move toward the vision discussed above. We hope that when these
systems reach maturity we will have the opportunity to share our
designs with the community.

\section{Conclusions}

There is a growing recognition that volume, velocity, and variety
require different models of computation and alternative processing platforms. We
certainly learned this lesson first hand in trying to deploy a
Hadoop-based solution for a problem it was ill-suited to solve. This
led us to implement the search assistance service twice. Although the
experience was instructive, we hope that future system designers can
benefit from our story and build the right solution {\it the first
  time}. Even better, it would be desirable to build a generic data
processing platform capable of handling both ``big data'' {\it and}
``fast data''.

\end{document}